\documentclass[preprint,showpacs,preprintnumbers,amsmath,amssymb]{revtex4}
\usepackage{graphicx}
\usepackage[usenames]{color}
\usepackage{amsmath,amssymb}
\input{epsf}
\usepackage{epsf}

\begin{document}

\title{The effect of curvature in thawing models}

\author{Sergio del Campo}
\email{sdelcamp@ucv.cl} \affiliation{ Instituto de F\'{\i}sica,
Pontificia Universidad Cat\'{o}lica de Valpara\'{\i}so, Avenida
Brasil 2950, Casilla 4059, Valpara\'{\i}so, Chile.}
\author{V\'ictor H. C\'ardenas}
\email{victor@dfa.uv.cl}

\affiliation{Departamento de F\'{\i}sica y Astronom\'ia,
Universidad de Valpara\'iso, Gran Breta\~na 1111, Valpara\'iso,
Chile}

\author{Ram\'on Herrera}
\email{ramon.herrera@ucv.cl} \affiliation{ Instituto de
F\'{\i}sica, Pontificia Universidad Cat\'{o}lica de
Valpara\'{\i}so ,  Avenida Brasil 2950, Casilla 4059,
Valpara\'{\i}so, Chile.}

\begin{abstract}
We study the  evolution of  spatial curvature for thawing class of
dark energy models. We examine the evolution of the equation of
state parameter, $w_\phi$, as a function of the scale factor $a$,
for the case in which the scalar field $\phi$ evolve in nearly
flat scalar potential. We show that all such models provide the
corresponding approximate analytical expressions for
$w_\phi(\Omega_\phi,\Omega_k )$ and $w_\phi(a)$. We present
observational constraints on these models.
\end{abstract}

\maketitle

\section{Introduction}

About a decade ago, current measurements of redshift and
luminosity-distance relations of Type Ia Supernovae
(SNe)\cite{S10} indicate that the expansion of the universe
presents an accelerated phase \cite{R98,P99}. In fact, the
astronomical measurements showed that Type Ia SNe at a redshift of
$z \sim 0.5$ were systematically fainted which could be attributed
to an acceleration of the universe caused by a non-zero vacuum
energy density. As this  shows  a result, that the pressure and
the energy density of the universe should violate the strong
energy condition, $\rho_\phi + 3\,p_\phi\,>\,0$, where $\rho_\phi$
and $p_\phi$ are energy density and pressure of some exotic,
unknown and unclustered matter component, dubbed dark
energy\cite{HT99} (see also Refs. \cite{T10,C09} for recent
reviews). A direct consequence of this,  is that the pressure must
be negative.

Various models of dark energy have been proposed so far. Perhaps,
the most traditional candidate to be considered is a non-vanishing
cosmological constant \cite{S03,T04}. Other possibilities are
quintessence \cite{C98,Z98}, k-essence \cite{Ch00,AP00,AP01},
phantom field \cite{C02,C03,H04}, holographic dark energy
\cite{Z07,W07}, etc. (see Ref.~\cite{D09} for model-independent
description of the properties of the dark energy and
Ref.~\cite{S09} for possible alternatives).

The first step toward understanding the property of dark energy is
to make clear whether it is a simple non-vanishing cosmological
constant or its genesis comes from other sources which dynamically
change in time. It is possible to distinguish between these two
possibilities by taking into account the evolution of the equation
of state parameter defined by $\omega_\phi\equiv
p_\phi/\rho_\phi$.

In what concern to the dynamical dark energy (or quintessence) its
physics is described by a scalar field, $\phi$, (quintessence
scalar field), with canonical momentum\cite{varios01}. One of the
main characteristic of the quintessence field is when it rolls the
self interacting potential curve. It will provide a negative
pressure if the potential curve is quite flat. In this way, the
quintessence scalar field evolves slowly enough to drive the
present cosmic acceleration.

Since the evolution of the quintessence scalar field my be
described by the change of the equation of state parameter
$w_\phi$, so that we could distinguish two possible situations:
the case in which $d\omega_\phi/d\phi<0$ and
$d\omega_\phi/d\phi>0$. The former case is referred as the
freezing and the later the thawing scenarios,
respectively\cite{CL05}(see also Ref. \cite{L06} for details).
While the observational data up to now are not discriminating in
the sense that we could not distinguish between a freezing or a
thawing phases by the variation of the equation of state
parameter, it is expect that will be able to do so with the next
decade high-precision astronomical observations.

On the other hand, in what concern to the curvature of the
universe, today we do not know precisely the geometry of the
universe, since we do not know the exact amount of matter present
in the Universe. Various tests of cosmological models, including
space–time geometry, galaxy peculiar velocities, structure
formation and very early universe descriptions (related to the
Guth´s inflationary universe model \cite{G81}) support a flat
universe scenario. However, by using the seven-year Wilkinson
Microwave Anisotropy Probe (WMAP) data combined with measurements
of Type Ia SNe and Baryon Acoustic Oscillations (BAO) in the
galaxy distribution, it was reported that the value for the
curvature density parameter, $\Omega_k
=-0.0057^{+0.0066}_{-0.0068}$ (68\% CL) represents a preferred
model, which is slightly closed\cite{KetAl10,LetAl10}.


In this paper we would like to study some of the consequences that
this slightly curvature may have on the evolution of the universe,
together with the the situation in which the thawing cosmological
evolution for the quintessence scalar field is invoked. The
outline of the paper goes as follow, in section II we present the
model to be study. Section III, deals with the fundamental field
equations which allow then and the  dynamical system. Finally, in
section IV we conclude with our finding.

\section{The model}

The Friedmann equation in which curvature is taken into account
becomes given by
\begin{equation}\label{H}
H^2 + {k \over a^2} = \frac{\rho}{3},
\end{equation}
where the Hubble parameter $H=\dot{a}/a$, with $a$ dot
representing a derivative with respect to the cosmological time,
$a$ is the scale factor, and the curvature parameter $k=0,+1,$ and
$-1$ represents flat, closed and open spatial section,
respectively. Here, we use units for which $8 \pi G = 1$. The
total energy density $\rho$ is given by $\rho=\rho_\phi+\rho_m$,
where $\rho_\phi$ and $\rho_m$
 are the energy density of dark energy and dark
matter, respectively. We will assume that these two components
are conserve separately, satisfying the continuity equations
\begin{equation}
\dot{\rho_m}+3H\rho_m=0,
\end{equation}
and
\begin{equation}
\dot{\rho_\phi}+3H(\rho_\phi+p_\phi)=\dot{\rho_\phi}+3H\rho_\phi(1+w_\phi)=0,\label{phi}
\end{equation}
where $w_\phi$ is the equation of state parameter introduced in
the introduction.

We assume that the dark  energy is modelled by a minimally-coupled
scalar field $\phi$, where the pressure and density of the scalar
field are given by
\begin{equation}
p_\phi = \frac{\dot \phi^2}{2} - V(\phi),
\end{equation}
and
\begin{equation}
\label{rhodense} \rho_\phi = \frac{\dot \phi^2}{2} + V(\phi),
\end{equation}
respectively. Here, $V(\phi)$ represents the effective potential
associated to the scalar field.

In term of the scalar field, Eq.(\ref{phi}) can be written as
\begin{equation}
\label{motionq} \ddot{\phi}+ 3H\dot{\phi} + V_{,\phi} =0.
\end{equation}
Equation (\ref{motionq}) indicates that the field rolls down the
hill in the scalar potential, $V(\phi)$, but its motion is damped
by a term proportional to $H$.

\section{Evolution  with curvature}

We will assume that the scalar field moves in a nearly flat scalar
potential, $V(\phi)$, quantitatively expressed as\cite{ScheSen08}
\begin{equation}
\label{slow1}
\left(\frac{1}{V} \frac{dV}{d\phi}\right)^2 \ll 1,
\end{equation}
and
\begin{equation}
\label{slow2}
\frac{1}{V}\frac{d^2 V}{d\phi^2} \ll 1.
\end{equation}
The  constraint given by Eq.(\ref{slow1}) ensures that $w_\phi\sim
-1$, meanwhile Eqs.(\ref{slow1}) and (\ref{slow2}) indicate that
$(1/V )(dV/d\phi)$ is nearly constant \cite{ScheSen08}. In the
nomenclature of Ref.\cite{CL05}, these are "thawing" models, i.e.
$d\omega_\phi/d \phi>0$.

From the Friedmann equation we get
$$
\Omega_k + \Omega_\phi + \Omega_m = 1,
$$
where the density parameters are $\Omega_\phi=\rho_\phi/(3H^2)$,
$\Omega_m=\rho_m/(3H^2)$ and $\Omega_k=-k/(aH)^2$, respectively.

Following, a similar technique developed  in Ref.\cite{ScheSen08},
Eqs.(\ref{H}) and (\ref{motionq}) can be expressed  in terms of
new variables $x$, $y$, $\lambda$, and $\Omega_k$, defined by
\begin{eqnarray}
\label{xevol}
x &=& \phi^\prime/\sqrt{6}, \\
y &=& \sqrt{V(\phi)/3H^2}, \\
\lambda &=& -V_{,\phi}/V,\\
\Omega_k &=& K/H^2,\label{Om}
\end{eqnarray}
where $K \equiv -ka^{-2}$ and a prime denote the derivative with
respect to $\ln a$, and $V_{,\,\phi}=d V(\phi)/d \phi$.

The density parameter $\Omega_\phi$ is expressed in terms of the
variables $x^2$ and $y^2$ in such a way that
\begin{equation}
\label{Om}
\Omega_\phi = x^2 + y^2,
\end{equation}
while, the equation of state parameter is given by
\begin{equation}
\label{gamma} \gamma \equiv 1+w_\phi = \frac{2x^2}{x^2 + y^2}.
\end{equation}

Eqs.(\ref{H}) and (\ref{motionq}) can be written in terms of the
new variables Eqs.(\ref{xevol})-(\ref{Om}), so that we get
\begin{eqnarray}
x^\prime &=& -3x + \lambda\sqrt{\frac{3}{2}}y^2 + \frac{3}{2}x\left[1 + x^2-y^2-\frac{\Omega_k}{3}\right],\label{1}\\
y^\prime &=& -\lambda\sqrt{\frac{3}{2}}xy + \frac{3}{2}y\left[1 + x^2-y^2-\frac{\Omega_k}{3}\right],\\
\lambda^\prime &=& - \sqrt{6} \lambda^2(\Gamma - 1) x,\\
\Omega_k^\prime &=& \Omega_k(1-\Omega_k+3[x^2-y^2])\label{2},
\end{eqnarray}
where
\begin{equation}
\label{Gamma}
\Gamma \equiv V \frac{d^2 V}{d\phi^2}/\left(\frac{dV}{d\phi} \right)^2.
\end{equation}
In the thawing model we have $w_\phi\sim -1$ and thus the $\gamma$
parameter satisfies $\gamma=1+w_\phi\ll 1$. Therefore, it is
useful to express Eqs.(\ref{1})- (\ref{2}) in terms of $\gamma$,
$\Omega_\phi$, $\Omega_k$ and $\lambda$, respectively. We obtain

\begin{eqnarray}
\label{gammaprime}
\gamma^\prime &=& -3\gamma(2-\gamma) + \lambda(2-\gamma)\sqrt{3 \gamma
\Omega_\phi},\\
\label{Omegaprime}
\Omega_\phi^\prime &=& 3(1-\gamma)\Omega_\phi(1-\Omega_\phi)-\Omega_\phi \Omega_k,\\
\label{zprime} \Omega_k^\prime &=& \Omega_k \left(1-\Omega_k +
3\Omega_{\phi}(\gamma -1) \right) \\
\label{lambda} \lambda^\prime
&=& - \sqrt{3}\lambda^2(\Gamma-1)\sqrt{\gamma \Omega_\phi}.
\end{eqnarray}

At this point we would like to stress two assumptions  that we are
considering: the first is
 $\gamma\ll 1$, which corresponds to $\omega_\phi\sim -1$,
as discussed previously. The second assumption we make is that the
scalar field begins with an initial value in a potential which is
nearly flat. In this way, following \cite{ScheSen08}, we assume
that $\lambda$ is approximately constant, so that
\begin{equation}
\label{lamb0} \lambda \approx \lambda_0 =
-(1/V)(dV/d\phi)\biggr|_{\phi =\phi_0}\ll\,1,
\end{equation}
where $\lambda_0$ is a small constant evaluated at $\phi=\phi_0$,
the initial value of the scalar field which corresponds to when it
stars to roll down the potential.

\begin{figure}[h]
\centering \leavevmode\epsfysize=7cm \epsfbox{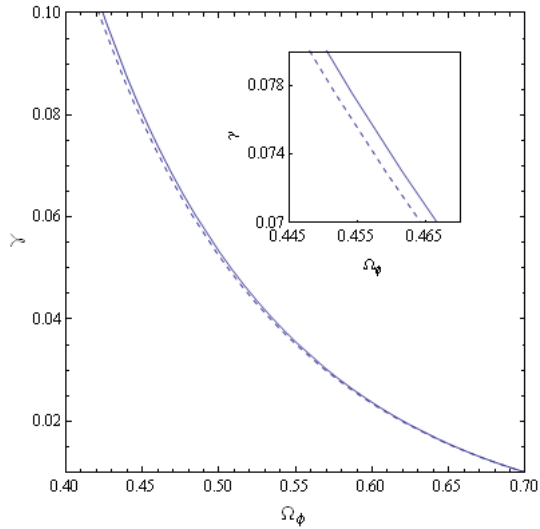}\\
\caption{\label{gvsfi} Numerical results for $\gamma$ as a
function of the fractional density parameter $\Omega_\phi$, for
nearly flat potentials. We have used the present values $\Omega_{k
0}=\pm 0.005$ and $\Omega_{\phi 0}=0.7$, with $\lambda_0=0.01$..}
\end{figure}

\begin{figure}[h]
\centering \leavevmode\epsfysize=7cm \epsfbox{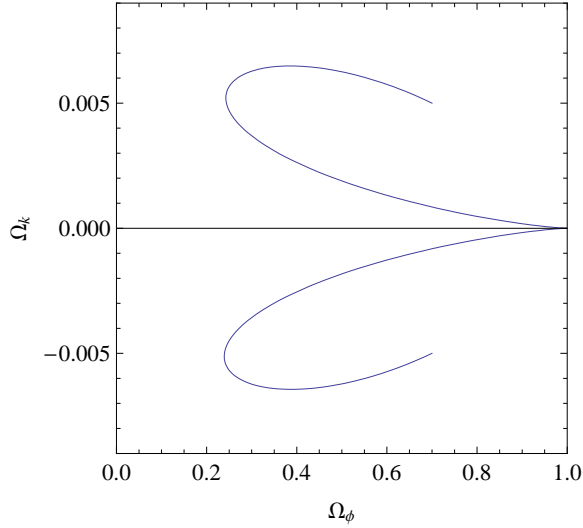}\\
\caption{\label{kvsfi} Numerical results for $\Omega_k$ as a
function of the fractional density parameter $\Omega_\phi$, for
nearly flat potentials.  We have used the present values
$\Omega_{k 0}=\pm 0.005$ and $\Omega_{\phi 0}=0.7$, with
$\lambda_0=0.01$.}
\end{figure}

\begin{figure}[h]
\centering \leavevmode\epsfysize=7cm \epsfbox{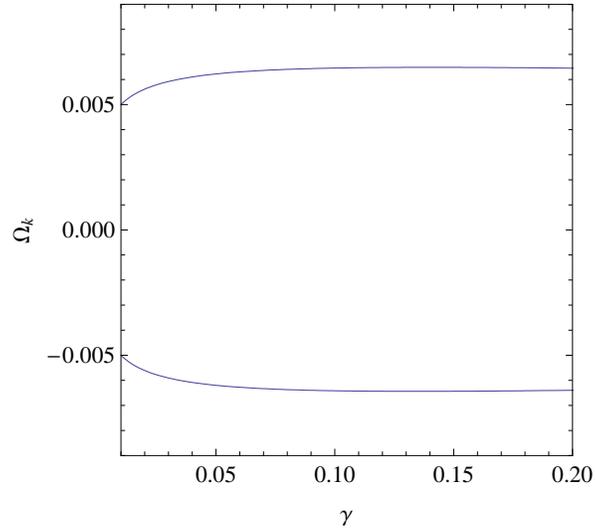}\\
\caption{\label{gamvsomk} Numerical results for $\Omega_k$ as a
function of  $\gamma$, for nearly flat potentials.  We have used
the present values $\Omega_{k 0}=\pm 0.005$ and $\Omega_{\phi
0}=0.7$, with $\lambda_0=0.01$.}
\end{figure}

Let us first to consider the evolution of the system using initial
values for the curvature  $\Omega_k = 0.005$. The result is shown in
figure \ref{gvsfi}. The same graph, but now with a value for the
curvature $\Omega_k = -0.005$ we get a similar plot with a small
difference when compared with the previous case.
The area in between the curves expands a continuous range of
values of the curvature parameter, well inside the current
observational constraints. In order to see this situation more
clearly, we plot in figure \ref{kvsfi} the projection of curves in
the $\Omega_k$-$\Omega_{\phi}$ plane. We observe that a large
region exist, even for small values $\Omega_k$.
As a complement, in figure \ref{gamvsomk} we show the degeneracies
in the variation of $\gamma$ with respect to the curvature. All
these figures are the result of a numerical integration of the
system of Eqs. (\ref{gammaprime}-\ref{lambda}) with
$\lambda_0=0.01$.

Motivated by the present value measured for the curvature
parameter we make the assumption that the curvature is a small
parameter, i.e. $\Omega_k \ll1$ along of the all story of the
evolution of the universe. From Eqs. (\ref{gammaprime}) and
(\ref{Omegaprime}) we can write
\begin{equation}\label{eqapp}
\frac{d\gamma}{d\Omega_{\phi}}=\frac{-6\gamma +2\lambda_0 \sqrt{3
\gamma \Omega_\phi}}{3\Omega_{\phi}(1-\Omega_{\phi})}+
\frac{\lambda_0 \Omega_k \sqrt{3 \gamma
\Omega_\phi}}{9\Omega_{\phi}(1-\Omega_{\phi})^2},
\end{equation}
where we have expanded and maintained the lowest order terms in
$\Omega_k$ and $\gamma$. Taking the boundary value $\gamma =0$ at
$\Omega_{\phi}=0$ (see Ref.\cite{ScheSen08}). The resulting
solution is
\begin{equation}\label{solEQ}
\gamma =
\frac{\lambda_0^2}{1296}\left[\frac{12}{\sqrt{\Omega_{\phi}}}
-\frac{\Omega_k
(\Omega_{\phi}+1)}{2(\Omega_{\phi}-1)\sqrt{\Omega_{\phi}}}+F(\Omega_k,\Omega_{\phi})
\right]^2,
\end{equation}
where
\begin{equation}\label{defF}
F(\Omega_k,\Omega_{\phi})=\left(\frac{\Omega_{\phi}-1}{\Omega_{\phi}}\right)
(12+{\Omega_k \over 2})\tanh^{-1} \sqrt{\Omega_{\phi}}.
\end{equation}
Note that in the limit of a flat universe, i.e., $\Omega_k
\rightarrow 0$, we recover the expression given in Ref.
\cite{ScheSen08}.

In the same way we can derive an approximate solution from Eqs.
(\ref{Omegaprime}) and (\ref{zprime}) under the same
approximations ($\gamma\ll1$ and $\Omega_k \ll1$)
\begin{equation}\label{ramon}
\frac{d\Omega_k}{d\Omega_{\phi}}=\frac{\Omega_k
(1-3\Omega_{\phi})}{3\Omega_{\phi}(1-\Omega_{\phi})},
\end{equation}
from which we get
\begin{equation}\label{solRamon}
\Omega_k = \Omega_{k0} \left[\frac{1-\Omega_\phi}{1-\Omega_{\phi
0}} \right]^{2/3}\left(\frac{\Omega_\phi}{\Omega_{\phi 0}}
\right)^{1/3}.
\end{equation}

From Eqs.(\ref{solEQ}) and (\ref{solRamon}) we get
\begin{equation}\label{solEQ2}
\gamma (\Omega_\phi) =
\frac{\lambda_0^2}{1296}\left[\frac{12}{\sqrt{\Omega_{\phi}}}
-\frac{\Omega_{k0} \left[\frac{1-\Omega_\phi}{1-\Omega_{\phi 0}}
\right]^{2/3}\left(\frac{\Omega_\phi}{\Omega_{\phi 0}}
\right)^{1/3}
(\Omega_{\phi}+1)}{2(\Omega_{\phi}-1)\sqrt{\Omega_{\phi}}}+G(\Omega_{\phi})
\right]^2,
\end{equation}
where  the function $G(\Omega_\phi)$ is given by
$$
G(\Omega_{\phi})=\left(\frac{\Omega_{\phi}-1}{\Omega_{\phi}}\right)
\left(12+{\Omega_{k0} \left[\frac{1-\Omega_\phi}{1-\Omega_{\phi
0}} \right]^{2/3}\left(\frac{\Omega_\phi}{\Omega_{\phi 0}}
\right)^{1/3} \over 2}\right)\tanh^{-1} \sqrt{\Omega_{\phi}}.
$$

We can use equation (\ref{Omegaprime}) to solve for $\Omega_\phi$
as a function of $a$ and thus determine $w_\phi(a)$.  Taking the
limit $\gamma \ll 1$ and $\Omega_k \ll 1$ in equation
(\ref{Omegaprime}) gives the following solution
\begin{equation}
\label{Oma} \Omega_\phi = \left[1 + \left(\Omega_{\phi 0}^{-1} - 1
\right)a^{-3} \right]^{-1},
\end{equation}
where $\Omega_{\phi 0}$ and $\Omega_{k 0}$ are the present values
of $\Omega_\phi$, $\Omega_k$, respectively, and we take $a=1$ at
present time. Combining Eq. (\ref{Oma}) with Eq. (\ref{solRamon})
we obtain an approximated solution for $\Omega_k (a)$. Then, with
the explicit expressions for $\Omega_\phi (a)$ and $\Omega_k (a)$,
and by using Eq. (\ref{solEQ}) we get explicitly the equation of
state parameter $\omega_\phi$, as a function of the scale factor
$a$, i.e. $w_\phi(a)$.

\begin{figure}[h]
\centering \leavevmode\epsfysize=5cm \epsfbox{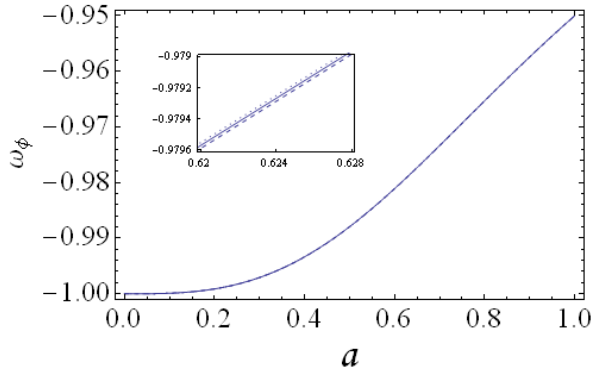}\\
\caption{\label{wvsa_curv}  Our anality results for the evolution
of the parameter $w_\phi$ as a function of the scale factor $a$
for nearly flat potentials. Here, we have taken the values
$w_0=-0.95$, $\Omega_{\phi 0}=0.7$, $\Omega_{k 0}=+0.005$ (dot
line), $\Omega_{k 0}=0$ (solid line) and $\Omega_{k 0}=-0.005$
(dash line), respectively. }
\end{figure}

In Fig.(\ref{wvsa_curv}) we show the dependence of the parameter
$w_\phi$ as a function of the scale factor $a$, for different values
of the curvature parameter $\Omega_k$ with $w_0=-0.95$ and
$\Omega_{\phi 0}=0.7$. Note that $w_\phi(a)$ is not sensible to the
value of $\Omega_{k}=0$ (see Ref.\cite{ScheSen08}).

We should mention that if we look for numerical solution to our
set of dynamical Eqs., in which  a scalar potential, such that
$V(\phi)\sim\phi^{2}$, $\phi^{-2}$, $\exp[-\phi]$, etc, is used,
we observe that there is no  much changes when them are compared
with that shown in Ref.\cite{ScheSen08}, where $\Omega_k=0$ was
taken into account.

\begin{figure}[h]
\centering \leavevmode\epsfysize=14cm \epsfbox{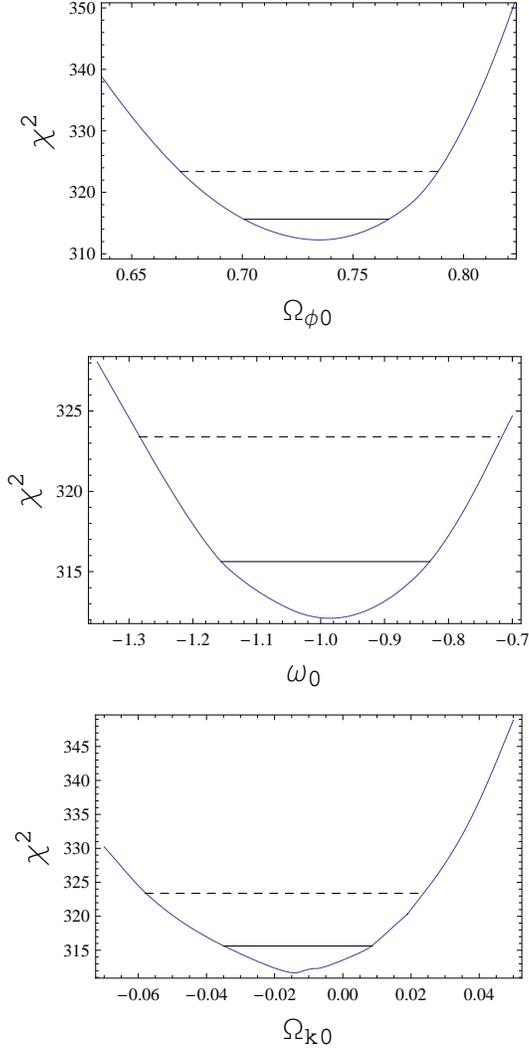}\\
\caption{\label{chi2} This panel shows the $\chi^2$ function
computed using the approximate analytical solution of $w_\phi(a)$.
The best fit parameters for $\Omega_{\phi 0}$,  $\omega_0$ and
$\Omega_{k0}$. The upper dashed line indicate the $99\%$
confidence range for each parameter and the continuous line below
indicate the $90\%$ of confidence range for each parameter.}
\end{figure}

Having an approximated expression for $w_\phi(a)$ we can use it to
perform a Bayesian analysis using SNIa observations, BAO distances
and CMB shift parameter. In this work, we use the Supernova
Cosmology Project Union sample \cite{SCP}, having $307$ SN
distributed over the range $0.015 < z < 1.551$. We fit the
(theoretical) distance modulus $\mu(z)_{th}$ defined by
\begin{equation}\label{muz}
\mu(z)_{th} = 5\log_{10}\left[ \frac{H_0 d_L(z)}{c}\right]+\mu_0,
\end{equation}
to the observational ones $\mu(z)_{obs}$. Here $H_0=100 h$km
s$^{-1}$ Mpc$^{-1}$ is the Hubble constant and the luminosity
distance is defined by $ d_L(z)=(1+z)r(z)$ where
\begin{equation}
r(z) = \frac{c}{H_0 \sqrt{ \left| \Omega_k \right|}} \text{Sinn}
\sqrt{ \left| \Omega_k \right|} \int_0^z \frac{dz'}{H(z')},
\end{equation}
and $ \mu_0  =  42.38 - 5\log_{10}h $. Sinn$(x)=\sin x, x, \sinh x$
for $\Omega_k<0$, $\Omega_k=0$, and $\Omega_k>0$ respectively. The
second major input for parameter determination comes from the baryon
acoustic oscillations (BAO) detected by Eisenstein et al.
\cite{ref:Eisenstein05}. In our work, we add the following term to
the $\chi^2$ of the model:
\begin{equation}
\chi^2_{BAO} = \left[ \frac{(A-A_{BAO})}{\sigma_A}\right]^{2},
\label{chiBAO}
\end{equation}
where $A$ is a distance parameter defined by
\begin{equation}
A = \frac{\sqrt{\Omega_m H_0^2}}{cz_{BAO}} \left[r^2(z_{BAO})
\frac{cz_{BAO}}{H(z_{BAO})})\right]^{1/3},
\end{equation}
and $A_{BAO}=0.469$, $\sigma_A = 0.017$, and $z_{BAO}=0.35$. The CMB
shift parameter $R$ is given by \cite{ref:Bond1997}
\begin{equation}
R(z_{\ast})=\sqrt{\Omega_m H^2_0}r(z_{\ast}).
\end{equation}
Here the redshift $z_{\ast}$ (the decoupling epoch of photons) is
obtained by using the fitting function \cite{Hu:1995uz}
\begin{equation}
z_{\ast}=1048\left[1+0.00124(\Omega_bh^2)^{-0.738}\right]\left[1+g_1(\Omega_m
h^2)^{g_2}\right],
\end{equation}
where the functions $g_1$ and $g_2$ are given as
\begin{eqnarray}
g_1&=&0.0783(\Omega_bh^2)^{-0.238}\left(1+ 39.5(\Omega_bh^2)^{0.763}\right)^{-1}, \\
g_2&=&0.560\left(1+ 21.1(\Omega_bh^2)^{1.81}\right)^{-1}.
\end{eqnarray}
The WMAP-7 year CMB data alone yields $R(z_{\ast})=1.726\pm0.018$
\cite{wmap7}. Defining the corresponding $\chi^2_{CMB}$ as
\begin{equation}
\chi^2_{CMB}=\frac{(R(z_{\ast})-1.726)^2}{0.018^2}\label{eq:chi2CMB},
\end{equation}
one can deduce constraints on $\Omega_{\phi 0}$, $\omega_0$ and
$\Omega_{k0}$. A joint analysis using SN+BAO+CMB leads to the best
fit values showed in Fig.\ref{chi2}, where we see the cross
section of the $\chi^2$ function in terms of the parameters
$\Omega_{\phi 0}$, $\omega_0$ and $\Omega_{k0}$. The two
horizontal lines indicate the $90\%$ and $99\%$ confidence range
for each parameter.

The analysis shows that considering thawing quintessence with an
explicit curvature term is consistent with observations. This is
exactly the conclusion of \cite{ScheSen08} for the flat case in
quintessence. However, as was demonstrated in \cite{SSS}, relaxing
the slow-roll assumption, the equation of state parameter for
different thawing potentials looks appreciably different. In the
following, we consider both quintessence and Tachyon field models,
and two scalar field potentials; $V=\phi$ and $V=\phi^{-2}$. In
figure \ref{fig6} we show the integration of the field equations for
current values of the curvature $\Omega_{k 0}= \pm 0.006$ and
$\Omega_{\phi 0}=0.72$ for all the models. The potential are
characterized by $\Gamma =0$ and $\Gamma=3/2$ respectively, along
the initial conditions $\lambda_{ini} \simeq 1$ (assuming that the
potential is not flat) and $\gamma_{ini} \simeq 0$ (the equation of
state parameter can vary from its freezing state ($w=-1$) until
today.)

\begin{figure}[h]
\centering \leavevmode\epsfysize=8cm \epsfbox{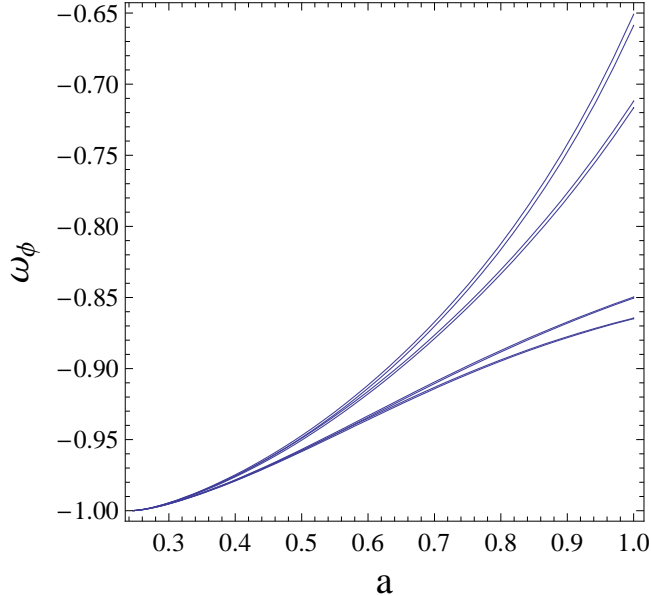}\\
\caption{\label{figure6} Here we show the numerical integration for
two scalar field models; quintessence and Tachyonic. We consider two
scalar field potentials $V=\phi$ and $V=\phi^{-2}$ and in each case
we use explicitly $\Omega_{k0}= \pm 0.006$ with a current value
$\Omega_{\phi 0}=0.72$. The upper two curves correspond to a
Tachyonic model with potential $V=\phi$. Although the values of the
curvature parameter are very small, the separation of the curves
increases with evolution and are appreciable here. The next couple
of curves correspond to a quintessence scalar field model with
potential $V=\phi$. The third set of two curves (which are closer
each other than the previous ones) correspond to a Tachyonic model
with potential $V=\phi^{-2}$. The bottom two curves correspond to a
quintessence model with $V=\phi^{-2}$. All these models have
$\Omega_{\phi 0}=0.72$.}
\end{figure}

\section{Conclusions}

In the present work we have studied the thawing dark energy
scenarios in which the effect of curvature was taking into
account.

We have plotted numerically  trajectories in the ($\gamma$,
$\Omega_\phi$), ($\Omega_k$, $\Omega_\phi$) and ($\Omega_k$,
$\gamma$) for a potential nearly flat.


We have shown that all such models converge to a common behavior
and we have find the corresponding approximate analytical
expressions for $\gamma(\Omega_\phi)$ given by Eq.(\ref{solEQ2})
and for $w_\phi(a)$ in the cases when $\gamma\ll 1$ and
$\Omega_k\ll 1$. Here, we noted that an analitical solution for
$w_\phi(a)$ is not very perceptible to the value of $\Omega_{k}
\neq 0$. A Bayesian analysis using SNIa data was performed to
constraint the best fit parameters using our analytic function,
$w_\phi(a)$. This analysis shows that current data does not rule
out the model. In this way, the motivation is to see whether one
can distinguish thawing dark energy models from $\Omega_k\neq 0$
models using this method.


\acknowledgments{ This work was funded by Comision Nacional de
Ciencias y Tecnolog\'{\i}a through FONDECYT Grants 1070306 (SdC)
and 1090613 (RH and SdC), and by DI-PUCV Grant 123787 (SdC) and
123703 (RH). }


\end{document}